\documentclass[aps,pre,preprint,groupedaddress,showpacs]{revtex4}
\begin{document}

\title{Dynamical model of steadily forced isotropic turbulence}

\author{Mohammad Mehrafarin}
\affiliation{Physics Department, Amirkabir University of Technology, Tehran 15914,
Iran}
\email{mehrafar@aut.ac.ir}

\date{\today}

\begin{abstract}
A dynamical model is proposed for isotropic turbulence driven by steady forcing that yields a viscosity independent dynamics for the small-scale (inertial) regime. This reproduces the Kolmogorov spectrum for the two-point velocity correlation function in the fully developed (stationary) stage, while predicting intermittency corrections for higher order moments. The model also yields a transient stage with a power-law time evolution. The crossover time to fully developed turbulence scales with the turbulent system size as $\sim L^{11/3}$. The physical origin of the transient behavior is explained.
\end{abstract}


\maketitle

\section{Introduction}
With regard to the problem of turbulence, the nonlinear inertial term in the Navier-Stokes equation is responsible for energy redistribution among the constituent length scales, as well as for the so-called sweeping effect, which couples any given scale with all the larger scales without redistribution of energy. Attempts to understand these nonlinear effects have led to different approaches. For instance, the nonperturbative nature of the sweeping interactions, which has been shown \cite {L'vov} to mask the classical scale-invariant theory of Kolmogorov \cite{Kolmogorov}, has led to the exploration of nonperturbative alternatives (e.g. \cite{McComb,Avellaneda,Eyink,Mou,L'vov2}). In forced systems, other alternatives such as the Burgers equation (e.g. \cite{Cheklov,Polyakov,HJ}), shell models (e.g. \cite{Jensen}) or their variants \cite{Grossmann} have been also explored. Some attention has been paid, as well, to the possible depletion of nonlinearity in parts of the real space \cite{Frisch}. Yet another approach, based on the idea of inventing an effective model for the nonlinearity, has in turn led to a large collection of phenomenological models \cite{Frisch2}. The model we shall introduce here for isotropic turbulence falls more or less into the last category.

Because isotropy everywhere implies homogeneity, by isotropic turbulence we mean, of course, homogeneous isotropic turbulence. In isotropic turbulence, in contrast with shear-flow turbulence, no mean shear stress can occur and consequently the mean motion of the fluid is uniform. Isotropy is a statistical property of the fluctuating part, ${\bf u}({\bf x},t)$, of the fluid's velocity field, i.e., a property of the velocity field with respect to an observer moving with the mean motion of the fluid. The isotropic field ${\bf u}({\bf x},t)$ satisfies
$$
<{\bf u}({\bf x},t)>=0, \ \ \ \ \ \ <u_i({\bf x},t) u_j({\bf x},t) >=\frac{1}{3} <{\bf u}^2({\bf x},t)> \delta_{ij}. 
$$
Assuming the flow incompressible, in the reference frame of the co-moving observer, the Navier-Stokes equation takes the form 
\begin{equation}
\partial_t {\bf u} + ({\bf u}.\nabla) {\bf u} = \nu \nabla^2 {\bf u}- \frac{1}{\rho} \nabla p+ {\bf f}, \label{2}
\end{equation} 
where $p({\bf x},t)$ is the fluctuating part of the pressure field ($<p>=0$) and ${\bf f}({\bf x},t)$ represents the fluctuating stirring force ($<{\bf f}>=0$). The pressure term is in effect a nonlinear term; it can be eliminated via the incompressibility condition in the standard manner (e.g. \cite{McComb}), at the price of introducing additional (and more complicated) nonlinearity in the equation. In view of the isotropy, it suffices to consider the dynamics of the velocity profile in a fixed direction. Thus, let $v({\bf x},t)$ and $\eta({\bf x},t)$, respectively, denote the fluctuating velocity and force components in an arbitrary direction. Equation (\ref{2}) in essence reads:
$$
\partial_t {v} = -\nabla . {\bf j}_{visc}+ \eta + nonlinear\ terms, 
$$
where ${\bf j}_{visc}=-\nu \nabla v$ is the viscous current and represents the momentum flux due to viscous stresses. The viscous current acts to oppose the spatial variations of $v$ and, thus, has a smoothing effect on its fluctuation-induced irregularities. Our approach is based on the idea of inventing an effective model for the nonlinearity by augmenting ${\bf j}_{visc}$ by a suitable `current' term ${\bf j}_{turb}$ according to
\begin{equation}
\partial_t {v} = -\nabla . ({\bf j}_{visc}+ {\bf j}_{turb})  + \eta. \label{3}
\end{equation}
Of course, ${\bf j}_{turb}$ must satisfy $<{\bf j}_{turb}>=0$ (up to a trivial constant). Physically, the augmenting inertial current ${\bf j}_{turb}$ is supposed to represent the momentum flux due to the collective action of turbulent fluctuations. The competition between the two currents ${\bf j}_{visc}$ and ${\bf j}_{turb}$ will be central to the turbulence problem. In equation (\ref{3}) the fluctuating fields $\eta$ and ${\bf j}_{turb}$ are to be deduced from  physical arguments and assumptions, rendering the basic equation of the model. The model we so present is new and will be shown to yield, through the competition of the two currents, a small-scale regime in which the dynamics is independent of viscosity. This regime corresponds to the inertial regime of the steadily forced isotropic turbulence. The equation associated with the inertial regime can be treated analytically by the dynamical renormalization group method to yield the scaling behavior of the two-point velocity correlation function. This reproduces the celebrated Kolmogorov scaling law \cite{Kolmogorov} for the fully developed (stationary) stage, while yielding the power-law time evolution $\sim t^{2/11}$ for the early transient stage. Furthermore, the model predicts intermittency behavior for higher order correlation functions in the fully developed stage. The crossover time from the transient to the stationary behavior scales as $\sim L^{11/3}$, where $L$ is the turbulent system size. The physical origin of the transient behavior will be explained. 

Interestingly, our model equation for the inertial regime (equation (\ref{6})) also appears in the context of surface growth phenomena \cite {BS} as a certain growth equation (see section III). While the similarity between the characteristics of surface growth models \cite{Bohr, Das}, in particular of this growth model \cite{Krug,Samanta}, and turbulence has been previously noticed, this work presents a first instance of derivation of such a model from pure dynamical considerations. This provides another perspective from which the model can be viewed. The fact that the Kolmogorov spectrum for isotropic turbulence has been derived more or less directly from the Navier-Stokes equation (e.g. \cite{McComb,Lundgren}), does not hinder the importance of other relevant model equations; after all the whole machinery of the Navier-Stokes  equation may not be required as far as the inertial range dynamics is  concerned. In fact, as is well known, in the zero viscosity (infinite Reynolds number) limit which is relevant to inertial range dynamics, the Navier-Stokes equation shows scaling symmetry for the velocity profile. This motivates us to put forward an analogy between interface growth phenomena and the inertial range dynamics. The analogy is based on the observation that the turbulent velocity profile may be viewed as an evolving interface that attains a self-affine form due to the roughness (width) generated by the collective action of turbulent fluctuations. The momentum flux due to these fluctuations, as shown in section II, can be obtained from dynamical considerations yielding the model equation.

\section{The model}
In view of the degree of control over the stirring forces exerted in an actual turbulent flow, it seems plausible to assume that in microscopic scale the forcing is not correlated in space and time. $\eta$ may thus be represented by an uncorrelated noise term, with the standard properties
$$
<\eta({\bf x},t)>=0, \ \ \ \ \ <\eta({\bf x},t)\eta({\bf x}^\prime, t^\prime)>=2D\ \delta^3({\bf x}-{\bf x}^\prime) \delta(t-t^\prime).
$$
Equation (\ref{3}), then, reads:
$$
\partial_t {v} = -\nabla . {\bf j} + \eta, 
$$
where ${\bf j}= {\bf j}_{visc}+{\bf j}_{turb}$ is the total current. At every point $v$ and its derivatives are independent variables that can develop (independent) irregularities through fluctuations and the total current acts as to oppose the resulting (independent) spatial variations by redistributing the irregularities. The viscous component ${\bf j}_{visc}$ of the current, as mentioned before, serves to iron out the spatial variations of the velocity field $v$ itself. The rest (irregularities of the derivative fields, most notably the energy dissipation rate) is to be compensated for by the effect of the inertial component ${\bf j}_{turb}$. We, therefore, hypothesize that the turbulent current has a similar smoothing effect on the spatial variations of the derivative fields, most notably the velocity gradient $\nabla v$. The variation of energy dissipation with scale, which is related to intermittency, will be, therefore, kept finite throughout. Then, the explicit form of ${\bf j}_{turb}$ may be deduced from the following dynamical requirements. 

(i) Temporal homogeneity: For turbulence driven by statistically steady forcing, the dynamics must be invariant under translations in time. This rules out an explicit time dependence of ${\bf j}_{turb}$. 

(ii) Spatial homogeneity: The dynamics must be invariant under translations in space. This excludes explicit ${\bf x}$-dependence of $\nabla .{\bf j}_{turb}$. 

(iii) Isotropy: The spatial derivatives of $\nabla v$ must appear isotropically. 

\noindent We can, thus, write the following general form:    
\begin{equation}
{\bf j}_{turb}=\lambda_1 (-\nabla) (\nabla v)^2+ \lambda_2 (-\nabla)^2 (\nabla v)+... \ , \label{4}
\end{equation}
where $\lambda$'s are positive constants. (The space (time) independence of $\lambda$'s follows from the invariance of $\nabla .{\bf j}_{turb}$ under space (time) translations, while their positivity is attributed to the smoothing effect of ${\bf j}_{turb}$ on the spatial variations of $\nabla v$.) Note that, because of homogeneity, each term in (\ref{4}) vanishes when averaged in compliance with $<{\bf j}_{turb}>=0$, as required by (\ref{3}).

We, thus, propose the following equation as our model of steadily forced isotropic turbulence:
\begin{equation}
\partial_t v=\nu \nabla^2 v+\lambda _1 \nabla^2 (\nabla v)^2-\lambda_2 \nabla^4 v +\eta. \label{5}
\end{equation}
This equation contains all the relevant terms describing the phenomenon. Scaling arguments indicate that terms arising from other first and second order derivatives of $\nabla v$ in ${\bf j}_{turb}$ (such as $(-\nabla) (\nabla v)^4, (-\nabla)^2 (\nabla v)^3,...$), as well as those arising from higher order derivatives, are all irrelevant to the scaling behavior of the equation and are, hence, disregardable.

Let us now introduce the size, $l$, of turbulent eddies, which is, by definition, the length scale over which the velocity field varies appreciably. The lower bound to $l$ is determined by the viscous dissipation length, $l_d$, and the upper bound by the turbulent system size, $L$, which is set by the stirring forces. The competition between the linear terms $\nabla^2 v$ and $\nabla^4 v$ in (\ref{5}), which represents the competition between viscous and inertial processes, generates a characteristic length scale $l_c=\surd (\lambda_2/ \nu)$ \cite{note}. For $l \ll l_c$, i.e. $\nu |\nabla^2 v| \ll \lambda_2 |\nabla^4 v|$, turbulent fluctuations dominate and the dynamics of the velocity profile can be represented by neglecting viscosity. This `small-scale' regime, thus, corresponds to the inertial regime of the steadily forced isotropic turbulence. The dynamics in this regime is, whence, governed by the nonlinear equation
\begin{equation}
\partial_t v=\lambda _1 \nabla^2 (\nabla v)^2-\lambda_2 \nabla^4 v +\eta. \label{6} 
\end{equation}
(In the `large-scale' regime, $l \gg l_c$, viscosity dominates and the dynamics will be, therefore, described by the linear Edwards-Wilkinson \cite{EW} equation. This regime is not relevant to inertial range dynamics.) If $l_c$ is large enough- comparable with the system size $L$- the inertial regime will involve the whole range $l_d \leq l \ll L$. This is generally expected to occur for large-scale/Reynolds-number flows. On the other hand, if $l_c$ is small, one should concentrate on very small eddy sizes ($\ll l_c$) in order to observe isotropic turbulence. This, of course, corresponds to the local isotropy hypothesis for laboratory-scale flows, which is corroborated by our model. The dynamics described by equation (\ref{6}) is independent of viscosity and is strongly influenced by the nonlinear term (as shown by scaling arguments). Viscosity plays role only in so far as to determine the extent of the inertial regime through $l_c$.

\section{The consequences}
The scaling behavior of the two-point correlation function $<v({\bf x},t) v({\bf x}^\prime,t)>$ in the inertial regime can be derived from a dynamical renormalization group analysis of equation (\ref{6}). This equation has also made appearance in surface growth problems \cite{BS}, where it was introduced by Lai and Das Sarma \cite{LS} to describe growth by the molecular-beam-epitaxy (MBE) process. As mentioned in the Introduction, the turbulent velocity profile may be viewed as an evolving (three-dimensional) hypersurface that attains a self-affine form due to the roughness generated by the collective action of turbulent fluctuations. The two-point velocity correlation function yields a measure of roughness of this self-affine hypersurface. The higher order correlation functions of the model were argued by Krug \cite{Krug} to display multiscaling (intermittency), the origin of which was traced back to the nonlinear term. This reconciles with the general understanding that intermittency sets in for the fluid because of the nonlinearity of the Navier-Stokes equation (e.g. \cite{L'vov,HJ}). Our model of steadily forced isotropic turbulence, thus, falls into the same universality class of scaling behavior as the MBE growth (albeit) on a (non-physical) three-dimensional substrate. We have the scaling form
\begin{equation}
<v({\bf 0},t) v({\bf x},t)>= x^{2\alpha} f(t x^{-z}), \label {7} 
\end{equation}
where $x=|{\bf x}|$, and the scaling function $f(y)$ satisfies $f(y \gg 1) \sim const.,\ f(y \ll 1) \sim y^{2\alpha/ z}$. In the language of interface problems, $\alpha$ and $z$ are the roughness and the dynamics exponents, respectively, which are calculated \cite{LS} to be $\alpha=1/3,\ z=11/3$ for dimension three. According to (\ref{7}), the velocity correlation begins to increase with time as $\sim t^{2\alpha/z}(\ =t^{2/11})$ but eventually reaches the stationary stage where it scales as $\sim x^{2\alpha}$. With $\alpha=1/3$, this is just the Kolmogorov's $2/3$ law for the fully developed (stationary) stage of isotropic turbulence. Of course, because Kolmogorov's theory deals only with the stationary stage of the fully developed turbulence, it does not at all comment on $z$. This exponent, which characterizes the time-dependent behavior of turbulence, is a dynamic feature of our model. As a corollary of the scaling form (\ref{7}), the transient time evolution and the stationary state stages can be characterized, respectively, by $t \ll \tau$ and $t\gg \tau$, where $\tau \sim L^z$. The crossover time, $\tau$, thus increases indefinitely with the turbulent system size implying that fully developed turbulence constitutes a finite-size effect. 

To understand the physical origin of the transient behavior, let us restate the above results in terms of the energy spectrum instead of the velocity correlation function. We define the Fourier transform pairs
\begin{eqnarray}
v({\bf x},t)= \frac {1}{\sqrt V} \sum_{\bf k} \tilde{v}({\bf k},t)\ e ^{i{\bf k}.{\bf x}} \nonumber \\ 
\tilde{v}({\bf k},t)= \frac {1}{\sqrt V} \int  v({\bf x},t)\ e ^{-i{\bf k}.{\bf x}}\ d^3 x, \nonumber  
\end{eqnarray}
where $V$ is the volume occupied by the fluid. One obtains, on using the homogeneity property, 
\begin{eqnarray}
<|\tilde{v}({\bf k},t)|^2>= \int <v({\bf 0},t) v({\bf x},t)> e ^{-i{\bf k}.{\bf x}}\ d^3 x \nonumber \\
<v({\bf 0},t) v({\bf x},t)>=\frac{1}{(2\pi)^3} \int <|\tilde{v}({\bf k},t)|^2> e ^{i{\bf k}.{\bf x}}\ d^3 k, \nonumber
\end{eqnarray}
having let $V \rightarrow \infty$ (as required for rigorous isotropy) {\it after} taking the averages. We thus find, using (\ref{7}), that 
$$
<|\tilde{v}({\bf k},t)|^2>= k^{-(2\alpha+3)} g(t k^z),
$$
where $k=|{\bf k}|$, and the scaling function $g(y)$ behaves asymptotically just like $f(y)$. The turbulent kinetic enrgy per unit mass of the fluid, namely,
$$
E= \frac{3}{2} <v^2>= \frac{3}{2}\frac{1}{(2\pi)^3} \int <|\tilde{v}({\bf k},t)|^2> d^3 k,
$$
whence yields the energy spectrum
\begin{equation}
{\cal E}(k,t)= \frac{3}{4\pi^2}\ k^{-(2\alpha+1)} g(t k^z), \label{8}
\end{equation}
for the inertial range $k_c \ll k \leq k_d$, where $k_c=2\pi/l_c$ and $k_d=2\pi/l_d$, of course. As implied by (\ref{8}), the energy per unit wave-number of mode $k$ begins to increase with time as $\frac{1}{k} t^{2\alpha/ z}$, eventually saturating with the stationary scaling form $k^{-(2\alpha +1)}$. Again, with $\alpha=1/3$, this is just the Kolmogorov's $-5/3$ spectrum. The transient rise in the energy spectrum can be understood as follows. The steady energy input from the stirring forces continually goes directly into the smallest $k$ modes that are associated with the largest-scale eddies being produced. Since initially only these small modes are present, energy dissipation, which is associated with the largest modes, does not take place. The absorbed energy, therefore, rises continually until it has had time to cascade to the largest modes in the fully developed stage. The energy build-up then stops because of dissipation, resulting in the stationary Kolmogorov spectrum. The transient behavior is, thus, attributed to the time lag between the energy injection at large scales and its dissipation at fine scales \cite{Pearson}. The larger the system size, the larger the range involved in the full cascade process and, hence, the larger the time lag is expected to be. This (together with the absence of dimensional parameters) explains the scaling behavior $\tau \sim L^z$ for the crossover time to the fully developed turbulence.


\begin{thebibliography}{widest-label}
\bibitem{note}The two inertial terms are of the same order, which implies $\lambda_2=\lambda_1 v_0$, where $v_0$ is a  typical velocity scale. Hence the nonlinear term can equally well enter the  estimate of $l_c$, yielding the same numerical value $l_c=\surd(\lambda_1 v_0/\nu)$.
\end{thebibliography}
\end{document}